\documentclass[useAMS,usenatbib,times,letter,amssymb]{mn2e}
\usepackage{epsfig,times,amssymb,amsmath,color,verbatim}
 
\usepackage{amsmath}
 
\usepackage{graphicx}% Include figure files 
\usepackage{bm}% bold math
\usepackage{natbib}

\def\ltsima{$\; \buildrel < \over \sim \;$}
\def\simlt{\lower.5ex\hbox{\ltsima}}
\def\gtsima{$\; \buildrel > \over \sim \;$}
\def\simgt{\lower.5ex\hbox{\gtsima}}

\def\[{\begin{equation}}
\def\]{\end{equation}}

\def\m@th{\mathsurround=0pt }
\def\eqalign#1{\null\,\vcenter{\openup1\jot \m@th
  \ialign{\strut\hfil$\displaystyle{##}$&$\displaystyle{{}##}$\hfil
  \crcr#1\crcr}}\,}

\title{The Size Distribution of Inhabited Planets}
\author[F. Simpson]{Fergus Simpson\thanks{fergus2@icc.ub.edu}
\\
ICC, University of Barcelona (UB-IEEC), Marti i Franques 1, 08028, Barcelona, Spain. \\ 
}

\date{\today}
\newcommand{\ud}{\mathrm{d}} 
   
\newcommand{\re}{r_{\oplus}}

\newcommand{\fevolve}{\eta(r)} % Don't use E, confused with expectation
\newcommand{\fixevolve}{\eta_0}

\newcommand{\surflux}{E_s} % Surface energy flux
 
\newcommand{\spm}{m_s} % Mass of species
\newcommand{\bioeff}{\eta_b} % biological efficiency. eta is standard efficiency symbol
\newcommand{\alienmass}{314} % Species mass (depending on final choice of priors) 
 
\newcommand{\MarsRad}{0.53 r_\oplus}
\newcommand{\EuropaRad}{0.25 r_\oplus}

\newcommand{\civchar}{\mathrm{\mathbf{C}}}
 \newcommand{\pc}{{\theta}}  % generic parameter 
  
\newcommand{\obs}{I}

\newcommand{\prim}{P} % Primitive life

\begin{document}

\maketitle

\begin{abstract} 
Earth-like planets are expected to provide the greatest opportunity for the detection of life beyond the Solar System. 
However our planet cannot be considered a fair sample, especially if intelligent life exists elsewhere.  Just as a person's country of origin is a biased sample among countries, so too their planet of origin may be a biased sample among planets. The magnitude of this effect can be substantial: over 98\% of the world's population live in a country larger than the median. In the context of a simple model where the mean population density is invariant to planet size, we infer that a given inhabited planet (such as our nearest neighbour) has a radius $r<1.2 \re$ (95\% confidence bound).  We show that this result is likely to hold not only for planets hosting advanced life, but also for those which harbour primitive life forms. 

Further inferences may be drawn for any variable which influences population size. For example, since population density is widely observed to decline with increasing body mass, we conclude that most intelligent species are expected to exceed $300$\,kg.
\end{abstract}

\section{Introduction} 
The discovery of extra-terrestrial life stands as one of the most ambitious objectives in modern scientific endeavour. Over one thousand distant planets have now been identified, spanning a broad spectrum of sizes and orbital configurations \citep{rice2014detection, pepe2011harps, bonfils2011harps, 2013Kepler}. Since many more await detection, only a small fraction can be subject to detailed follow-up investigations. We must therefore identify those deemed most likely to host life. At present the Earth is our only example of an inhabited planet, so its physical characteristics appear to provide a natural template for finding life elsewhere. However, as we shall see, selection effects may have biased our observational sample.    
% PLATO paper
 % Our existence has also been used to consider the fate of the human species, such as the so-called Doomsday argument \cite{carter1983anthropic, gott1993implications}. 
 % HARPS \cite{pepe2011harps} and HARPS-N \cite{cosentino2012harps}
 % from missions such as Kepler \cite{koch2010kepler} and PLATO \cite{2014ExA...tmp...41R},

It has often been postulated that our existence in the Universe could explain the magnitudes of various quantities in fundamental physics, such as the fine-structure constant, the cosmological constant  and primordial density perturbations \citep{1974IAUS...63..291C, carter1983anthropic, 1987PhRvL..59.2607W, 1995MNRAS.274L..73E, PhysRevD.52.3365, 2007PeacockAnthropic, 2014JimenezAnthropic, tegmark1998cosmic, tegmark2006dimensionless}.  If an ensemble of cosmological conditions exists, we should expect to observe those which permit the emergence of life. Or more specifically, those which maximise the abundance of life.  While this work will follow a similar line of reasoning, our approach differs from most in that there is no requirement for an ensemble of universes to exist. 

% Key goals
The physical characteristics of the Earth are considered to be the gold standard for habitability \citep[see e.g.][]{kasting1993habitable, schulze2011two}. 
However, for any non-singular distribution of population sizes, typical beings do not live within typical (median) populations. This is a statistical truism, yet it ensures that a violation of the mediocrity principle is inevitable. That we should expect to be a part of a large civilisation has been stated previously by \citet{gott1993implications}. Here we elaborate on this by considering the bias induced on observables which influence population size, such as the size of the host planet. We shall also generalise this result to incorporate the broader collection of planets which harbour primitive life forms.  

\section{Population Bias}

In the absence of any extra information, the probability of belonging to a particular group is proportional to the total membership of that group. This selection effect is apparent in numerous personal characteristics: your blood type, your class size at school, your employer, and your geographic location. It seems unremarkable to note that you are more likely to have a common blood type than a rare one, or that you are more likely to be living in China than the Cayman Islands. But if mankind's colonisation had spread beyond multiple continents to include another planet, the likelihood of belonging to this second planet must be weighted in the same manner. Of particular importance to this work is the following supposition: if the second colony had not arrived from Earth but instead evolved independently, it appears difficult to justify why our calculation should no longer hold.

Throughout this work we shall assume that the Universe hosts an ensemble of planets with advanced civilisations. We define an advanced civilisation as a  population of observers which has (a) colonised most of its host planet/moon and (b) developed sufficient intelligence to contemplate the existence of other inhabited planets. It is helpful to express the total number of observers $N$ produced by a given civilisation in the form

\[ \label{eq:observers}
N =  \frac{x L}{R} \, .
\]
Here $L$ is the longevity of the civilisation, over which time it sustains a mean population size $x$.  The mean lifespan averaged over all individuals is denoted by $R$. No assumption is made regarding the temporal variation of either the population size or birth rate.  If the ensemble of civilisations is suitably large we may consider a continuous probability distribution for the set of civilisation characteristics $\civchar \equiv \{ x, L, R , \ldots \}$. This encompasses a broad range of variables, including the physical characteristics of the host planet, such as its radius, density, and atmospheric properties. 

We wish to determine how a civilisation selected at random, such as our nearest neighbour, compares to our own. 
Our nearest neighbour is drawn from the ensemble of civilisations which exist today, $p( \civchar | T)$, at a fixed time $T$. The Earth is drawn from a different distribution, $p( \civchar | I)$, the ensemble of civilisations as sampled by a random observer,  $I$. 
% There are two distinct distributions of interest: the first is the ensemble of civilisations which would be found if we performed a thorough census of the cosmos. This corresponds to those civilisations which exist today, at a fixed time $T$.  
Using Bayes' theorem,  $p( \civchar | T) \propto L   p(\civchar)$, since $p(T | \civchar)  \propto L $, demonstrating that we are more likely to co-exist with long lived civilisations. Similarly, we find $p( \civchar | I)  \propto  N  \, p( \civchar )$.  Substituting  (\ref{eq:observers}), these two distributions are then related by $p( \civchar | I)  \propto \frac{x}{R} \, p( \civchar | T)$. In order to determine the probability distribution for a single element $ \theta \in \civchar$, we marginalise over all other parameters, yielding

 \begin{figure}
\includegraphics[width=80mm]{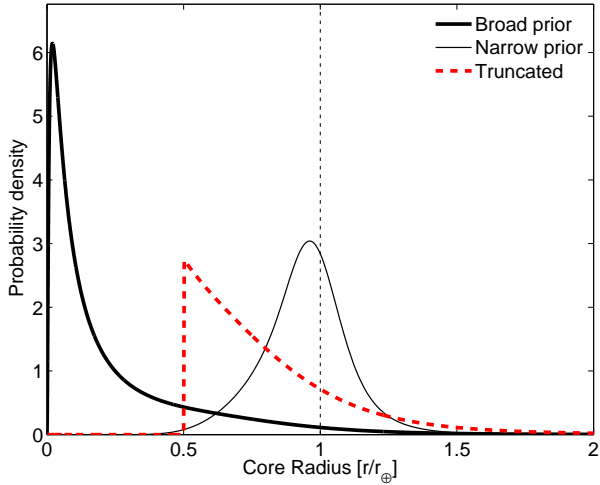}
\caption{Constraints on the radius $r$ of an inhabited planet, based on a constant mean population density, and after marginalising over the $\mu$ and $\sigma$ parameters of the lognormal distribution. The thick and thin solid lines correspond to the higher and lower set of $\sigma$ values. The dashed line illustrates the effect of imposing the condition $r > 0.5 r_\oplus$, as may be the requirement for an atmospheric water cycle. A common feature is the manner in which super-Earths are disfavoured, regardless of the choice of prior. Broad and narrow lines yield $95\%$ confidence bounds of $ r < 0.9  r_\oplus$ and $ r < 1.2 r_\oplus$  respectively, while the dashed line sets $ r < 1.4  r_\oplus$.  For reference, the radii of Europa and Mars are $\EuropaRad$   and $\MarsRad$ respectively.
\label{fig:planet_likeli}}
\end{figure} 
 
\[\label{eq:planetbias}  % Now need to average, unlike before
p( \pc | I)   \propto p( \pc | T) E\left(\frac{x}{R} \biggr| \pc, T\right) , 
\]
where the conditional expectation is defined as follows
\[ \label{eq:expectation}
 E\left(\frac{x}{R} \biggr| \pc, T\right)  \equiv  \iint \frac{x}{R} p(x,R | \pc, T)  \ud x \,  \ud R \, . 
\]
This is a general result, which makes no assumptions regarding the functional form of $p(x)$. If the expectation $E\left(\frac{x}{R} | \pc, T\right) $  is sensitive to the value of $\pc$, then $p( \pc | I)$ will differ from $p( \pc | T)$. In other words, provided the mean population of advanced civilisations is correlated with \emph{any} planetary characteristic, then the Earth is a biased sample among inhabited planets. This is the central result of this work. We now consider the particular case where  the generic parameter $\theta$ is taken to be the planet radius $r$.
 
 \section{Planetary Selection Effects}

The distribution of planetary radii has now been measured down to terrestrial sizes. \citet{2015SilburtPlanetRadii} found that across the range $1< r/r_\oplus < 4$, the distribution $p(r)$ appears to be approximately constant in log space. This was inferred from the radii of planets discovered by the Kepler spacecraft, taking into account the various observational selection effects.  However the distribution of planetary radii which host life remains highly uncertain. 
 
We seek to estimate the radius of another inhabited planet, such as our nearest neighbour, which is drawn from $p(r| T)$. The Earth's radius is our data point, representing a single sample drawn from $p(r| \obs )$. Ordinarily a solitary sample is insufficient to draw any conclusions about the range of the parent distribution, since an estimate of the variance requires at least two points. However, when biased, a single sample can be sufficient to impose a useful bound. For example, rolling a die once does not reveal any information on the values of the other faces. But if the die is loaded, such that each face's probability is proportional to its value, then we can infer from a single throw that the other faces have values lower or very similar to the observed face. 
 
 It would be unreasonable to expect the mean population of inhabited planets to remain constant for different radii. The available surface area increases, as does the total amount of available energy.   For small perturbations in planetary radius we should expect the mean population density to remain approximately constant. This suggests the expected population size rises as $E(x|r, T) \propto r^2$.  This scaling relation implies that the mean observed radius $\mu_I$ is related to the mean  radius $\mu$ by
  \[ \label{eq:r_bias}
\mu_I = \mu \left( 1 +  \frac{2  \sigma^2 + \frac{\mu_3}{\mu}}{\sigma^2 + \mu^2} \right) \, ,
\]
where $ \sigma^2$ and  $\mu_3$ are the variance and third moment of the distribution of inhabited radii. 
For \emph{any} non-singular distribution $(\sigma^2>0)$, observers should expect to find themselves on a larger planet than if the planet had been selected at random from the ensemble of inhabited planets. The broader the distribution of radii, the stronger the observational bias. 
 
To progress quantitively, we marginalise over a family of possible distributions for $p(r| T)$. The functional form is not particularly important, although it should ensure $p(r| T)$ falls to zero for very small and very large values of $r$. Here we adopt a Gaussian distribution in log space, $\ln\mathcal{N}(\mu_{r}, \sigma_r^2)$, marignalising over its mean and variance.  The full posterior likelihood $p(r|D, T)$ of inhabited planetary radii given our data $D$ (the radius of the Earth) is then
\[
p(r | D, T) \propto  \iint p(D | \mu_r, \sigma_r) p(r | \mu_r, \sigma_r)   \pi(\mu_r, \sigma_r^2) \ud \mu_r \ud \sigma_r^2 \, .
\]
where we adopt a reference prior $\pi(\mu_r, \sigma_r^2) \propto \sigma_r^{-2}$.  

Figure \ref{fig:planet_likeli} shows the likelihoods resulting from two different priors on $\sigma_r$. The thin solid line corresponds to a range of narrow distributions $0.05 < \sigma_r <  0.2$. For these low values of $\sigma_r$, the total distribution is narrow so radii far from the Earth are always disfavoured, and we find $r \! < \! 1.2  \re$ ($95\%$ confidence bound).  The thick solid line spans $0.2 < \sigma_r < 0.8$, setting a tighter bound $r \!< \! 0.9  \re$. At these larger values of $\sigma_r$, there is a strong selection effect at work, as quantified by (\ref{eq:r_bias}). This favours small planets, leaving the Earth to appear highly atypical.  More extreme values of $\sigma_r $, either larger or smaller than those considered here, only serve to tighten these bounds on $r$ further. The extent to which the Earth overestimates the radii of other inhabited planets strongly depends on the choice of prior. However the conclusion that larger radii are disfavoured is robust to the choice of prior on $\sigma_r$. 
 
Thus far we have deliberately neglected criteria which are conventionally assumed to influence habitability. If the emergence of life needs water, we can set a more stringent limit on the range of habitable radii. The smallest planets are not expected to be able to retain a thick enough atmosphere to sustain a water cycle.  Truncating the posterior likelihood generated from the broad prior such that $p(r \! < \! 0.5 \re) = 0$ leads to a modest amplification of the likelihood at larger radii, as shown by the dashed line.  In this case we find the $95\%$ confidence bound to be $ r < 1.4 \re$.  

\begin{figure}
\includegraphics[width=80mm]{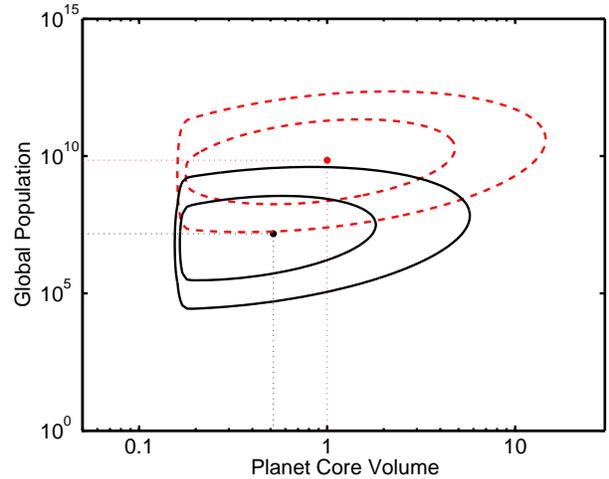}
\caption{An illustration of the biases in both population and planet size which can be induced when sampling on a \emph{per-individual} basis rather than a \emph{per-population} basis.  The solid contours represent where 68\% and 95\% of civilisations would be found in this model. The dashed contours represent the distribution as sampled by an individual.  In each case the median values are highlighted by dotted lines. In this example we calibrate the \emph{per-individual} median to that of the Earth's volume and estimated population of seven billion, yielding a typical civilisation of 15 million inhabiting a host planet approximately half the size of the Earth.    \label{fig:planet_contour}}
  \end{figure}

These results involve marginalising over a range of possible distributions. Figure \ref{fig:planet_contour} shows the specific case where $\sigma_r = 0.4$ and $\mu_r = 1$. The contours reflect the  $68\%$ and $95\%$ confidence limits for the planet-sampled (solid) and observer-sampled (dashed) distributions.
At the smallest radii the distributions are truncated due to atmospheric mass loss. The planet inhabited by the typical observer is over five times larger than the median planet size. The magnitude of this bias is predominantly determined by the breadth of the distribution, $\sigma_r$. 

Note that we have only estimated the relative abundance of inhabited planets, and make no statement regarding their overall prevalence in the Universe. These results are insensitive to the distribution of population sizes (it need not be lognormal), to the variables which appear in the Drake equation \citep{drake1992anyone}, and to the numerous variables which influence population size, provided they remain uncorrelated with planet size.

\section{ The Distribution of Primitive Life}

Thus far our calculation applies to planets harbouring advanced life forms. How does this relate to the question of where primitive life resides? Primitive life-forms are a pre-requisite for advanced life, and so their host planets must trace at least the same volume of parameter space as those of intelligent species.  The probability density function for a planet hosting primitive life, $p(r|\prim)$, may be expressed as 

\[ \label{eq:evolve}
p(r|\prim) \propto  \frac{1}{ \fevolve } p(r|T) \, ,
\]
where  $\fevolve$ is the proportion of life-bearing planets which host advanced life, as a function of radius $r$. If this function is constant, $\fevolve = \fixevolve$, then the two distributions are identical.  However larger biospheres host a wider range of species and environments, and a greater number of individual life-forms. For these reasons it appears likely that larger planets possess a greater probability of producing advanced life from a primitive state.  If we conservatively adopt $\fevolve \propto r$, the upper bound derived from the broad prior in Figure \ref{fig:planet_likeli} shrinks to yield $r<0.2 \re $. The truncated likelihood, as illustrated by the dashed line in Figure \ref{fig:planet_likeli}, tightens slightly to  $r<1.2 \re $ ($95\%$ confidence bound).
 
Even if mankind represents the only form of advanced life in the Universe, the fact that $\fevolve$ is likely to increase for larger values of $r$ suggests that the Earth is probably larger than most life-bearing planets.

\section{Characteristics of Advanced Species}

There are likely to be a number of other variables, aside from the size of the host planet, which are subject to selection bias due to their influence on population size. For example, species with a lower body mass are able to sustain a higher population density. This is a trend which has been extensively observed throughout the animal kingdom \citep{damuth1981population, damuth1987interspecific, loeuille2006evolution}. 
One proposed mechanism originates from Kleiber's law, the scaling relation linking the basal metabolic rate (BMR) to the body mass  $\spm$, $\mathrm{BMR} \propto \spm^{3/4}$ \citep{kleiber1932body, agutter2004metabolic}.  Given the finite energy resources available, population density drops as the individual's energy demand rises. Ants and termites vastly outnumber humans due to their small size.  If mankind tried to match their population, our total metabolic demand would exceed the entire solar flux incident upon our planet. 

We adopt a scaling relation between mean population and body mass given by $E(x |  \spm, T) \propto \spm^{-3/4}$.  Some variation in the value of this exponent has been observed  \citep{loeuille2006evolution}, however these were generally found to be steeper relationships, thus our model is a conservative one.  It is also the case that larger animals live longer \citep{speakman2005body, hulbert2007life}, suggesting $R \propto \spm^{1/4}$. Therefore we estimate that the conditional expectation defined in $(\ref{eq:expectation})$ is inversely proportional to body mass, $E(x/R |  \spm, T) \propto \spm^{-1}$.

The distribution of body masses among species on Earth can be well described by a lognormal \citep{greenwood1996relations}. Again we marginalise over the two lognormal parameters with a reference prior $\pi(\mu_m, \sigma_m^2) \propto \sigma_m^{-2}$. It would be extremely surprising if the diversity in body mass among extra-terrestrial species is lower than that amongst a small group of closely related species on Earth.  There are seven species of great ape, spanning gorillas,  orangutans, humans, and chimpanzees. Their body masses exhibit an inter-species standard deviation in log space of $\sigma_m \simeq 0.5$, which serves as our lower bound. For an upper bound we adopt $\sigma_m = 3$, so as not to greatly exceed the terrestrial variance. 

\begin{figure}
\includegraphics[width=80mm]{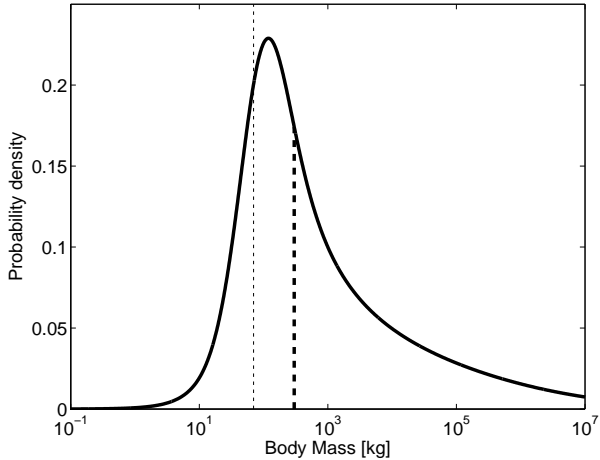}
\caption{ One-dimensional constraint on the body mass of an intelligent species after marginalising over the two lognormal parameters.  The thin dashed line represents our data point, at 70\,kg, while the thick dashed line represents the median mass, at $\alienmass$kg. This offset is predominantly   due to the expected fall in population size with increasing body mass.  For reference the adult African Elephant is approximately 6,000\,kg, while the heaviest dinosaurs such as Argentinasaurus were thought to be approximately $10^5$\,kg \citep{sellers2013march}. \label{fig:species_likeli}}  
\end{figure}

Figure \ref{fig:species_likeli} illustrates the probability density function for body mass, as derived from our single data point of $70$\,kg. The median body mass is found to be $\alienmass$\,kg, while the $95\%$ lower bound is given by $m_s > 25$\,kg. It is likely some correlation exists between planet size and median body mass, due to the influence of surface gravity, but we do not attempt to model this here.

 % STELLAR FLUX
In order to further explore the range of variables which influence the mean population size $x$, and are therefore susceptible to selection bias, we can decompose the contributing factors as follows

\[ \label{eq:x}
x = \bioeff A \frac{\surflux}{E_i}  \, ,
\] 
where the physical factors are the available area $A$, and  the energy flux density at the surface $\surflux$. The biological terms are the energy demand of the individual  $E_i$, and the energy efficiency of the species $\bioeff$. 
If the highest populations are resource limited, this may suggest we are receiving an unusually high radiation flux $\surflux$. This could arise either from being relatively close to our host star, or by possessing a lower atmospheric opacity.  In red dwarf systems most of the incident radiation is in the infrared which could lead to considerably lower values for $\surflux$.
%  For tidally locked planets, the habitable area $A$ may be significantly reduced. 

\section{Conclusions}
% POPULATION BIAS
On purely statistical grounds, any given individual should expect to be part of a larger group, not an ordinary group. 
Therefore, unless we are alone in the Universe, our planet is likely to be one which produces observers at a higher rate than most other inhabited planets. This may be accomplished by having a relatively large population, in combination with a low individual life expectancy. 
% PLANET SIZE
Any variable which correlates with population size will also be subject to observational bias.  By adopting a simple model where the mean population density is constant, we find that an inhabited planet selected at random has a radius $r<1.2 \re$ (95\% confidence bound).   

Our conclusions are not restricted to the search for intelligent life.  Provided the emergence of advanced life from primitive forms is not more likely on smaller planets, then the upper bound $r<1.2 \re$ also applies to a planet hosting life of any kind. 

Two distinct methods are currently being pursued for finding life on exoplanets: biomarkers within the atmospheric spectra of exoplanets \citep{hedelt2013spectral}, and somewhat more speculatively, the reception of radio signals from advanced civilisations \citep{tarter2011first}. In each case the signal is extremely challenging to detect, and it is therefore vital to correctly prioritise the strongest candidates. Larger planets and larger populations might provide stronger signals. However since they are expected to be relatively scarce this gain may be offset by their greater distance from the Earth.  

% SPECIES MASS
Aside from the size of the host planet, there are a number of other variables which influence population size, and these are also subject to observational bias. Throughout the animal kingdom, species which are physically larger invariably possess a lower population density, possibly due to their enhanced energy demands. As a result, we should expect humans to be physically smaller than most other advanced species. By marginalising over a feasible range of standard deviations, we conclude that most species are expected to exceed $300$\,kg in body mass. The median body mass is similar to that of a polar bear.

 % INTELLIGENCE
While larger species possess larger brains, the correlation between brain size and intelligence is weak. Higher intelligence enables the development of technologies which can sustain  larger population sizes. However it could also enable the longevity of individuals to increase substantially, thereby pushing the selection bias in the opposite direction. The net effect remains unclear. 
 
% Mention Fermi paradox?!
 The degree to which mankind and the Earth are atypical hinges on the level of diversity among advanced life forms. As we have repeatedly learned from the discoveries of distant planets, and the exploration of life on our own, nature is invariably more diverse than we anticipate, not less. 

\noindent{\bf Acknowledgements}\\
The author would like to thank R. Jimenez, L. Verde, J. Peacock, M. Rees, K.Rice, A. Lewis, J. Wee, S. Ho and D. Forgan for helpful discussions.  The author acknowledges support by the European Research Council under the European Community's Seventh Framework Programme FP7-IDEAS-Phys.LSS 240117. It should be noted that this article has been refused a listing on astro-ph for several months. It is the author's firm belief that new scientific ideas ought to be subject to skepticism, but not censorship. 

\setlength{\bibhang}{2.0em}
\setlength{\labelwidth}{0.0em}
\bibliographystyle{mn2e}
\bibliography{anthropic_bib}

\begin{thebibliography}{29}
\expandafter\ifx\csname natexlab\endcsname\relax\def\natexlab#1{#1}\fi

\bibitem[{Agutter \& Wheatley(2004)}]{agutter2004metabolic}
Agutter P.~S., Wheatley D.~N., 2004, Theoretical Biology and Medical Modelling,
  1, 13

\bibitem[{{Bonfils} {et~al}\mbox{.}(2013){Bonfils}, {Delfosse}, {Udry},
  {Forveille}, {Mayor}, {Perrier}, {Bouchy}, {Gillon}, {Lovis}, {Pepe},
  {Queloz}, {Santos}, {S{\'e}gransan}, \& {Bertaux}}]{bonfils2011harps}
{Bonfils} X. {et~al.}, 2013, \aap, 549, A109

\bibitem[{{Carter}(1974)}]{1974IAUS...63..291C}
{Carter} B., 1974, in IAU Symposium, Vol.~63, Confrontation of Cosmological
  Theories with Observational Data, {Longair} M.~S., ed., pp. 291--298

\bibitem[{Carter \& McCrea(1983)}]{carter1983anthropic}
Carter B., McCrea W.~H., 1983, Philosophical Transactions of the Royal Society
  of London. Series A, Mathematical and Physical Sciences, 310, 347

\bibitem[{Damuth(1981)}]{damuth1981population}
Damuth J., 1981, Nature, 290, 699

\bibitem[{Damuth(1987)}]{damuth1987interspecific}
Damuth J., 1987, Biological Journal of the Linnean Society, 31, 193

\bibitem[{Drake \& Sobel(1992)}]{drake1992anyone}
Drake F., Sobel D., 1992, Is anyone out there? Delacorte Press

\bibitem[{{Efstathiou}(1995)}]{1995MNRAS.274L..73E}
{Efstathiou} G., 1995, \mnras, 274, L73

\bibitem[{Gott(1993)}]{gott1993implications}
Gott J.~R., 1993, Nature, 363, 315

\bibitem[{Greenwood {et~al}\mbox{.}(1996)Greenwood, Gregory, Harris, Morris, \&
  Yalden}]{greenwood1996relations}
Greenwood J.~J., Gregory R.~D., Harris S., Morris P.~A., Yalden D.~W., 1996,
  Philosophical Transactions of the Royal Society of London. Series B:
  Biological Sciences, 351, 265

\bibitem[{{Hedelt} {et~al}\mbox{.}(2013){Hedelt}, {von Paris}, {Godolt},
  {Gebauer}, {Grenfell}, {Rauer}, {Schreier}, {Selsis}, \&
  {Trautmann}}]{hedelt2013spectral}
{Hedelt} P. {et~al.}, 2013, \aap, 553, A9

\bibitem[{Hulbert {et~al}\mbox{.}(2007)Hulbert, Pamplona, Buffenstein, \&
  Buttemer}]{hulbert2007life}
Hulbert A., Pamplona R., Buffenstein R., Buttemer W., 2007, Physiological
  Reviews, 87, 1175

\bibitem[{Kasting {et~al}\mbox{.}(1993)Kasting, Whitmire, \&
  Reynolds}]{kasting1993habitable}
Kasting J.~F., Whitmire D.~P., Reynolds R.~T., 1993, Icarus, 101, 108

\bibitem[{Kleiber(1932)}]{kleiber1932body}
Kleiber M., 1932, ENE, 1, E9

\bibitem[{Loeuille \& Loreau(2006)}]{loeuille2006evolution}
Loeuille N., Loreau M., 2006, Ecology Letters, 9, 171

\bibitem[{{Peacock}(2007)}]{2007PeacockAnthropic}
{Peacock} J.~A., 2007, \mnras, 379, 1067

\bibitem[{{Pepe} {et~al}\mbox{.}(2011){Pepe}, {Lovis}, {S{\'e}gransan}, {Benz},
  {Bouchy}, {Dumusque}, {Mayor}, {Queloz}, {Santos}, \& {Udry}}]{pepe2011harps}
{Pepe} F. {et~al.}, 2011, \aap, 534, A58

\bibitem[{{Petigura} {et~al}\mbox{.}(2013){Petigura}, {Howard}, \&
  {Marcy}}]{2013Kepler}
{Petigura} E.~A., {Howard} A.~W., {Marcy} G.~W., 2013, Proceedings of the
  National Academy of Science, 110, 19273

\bibitem[{{Piran} \& {Jimenez}(2014)}]{2014JimenezAnthropic}
{Piran} T., {Jimenez} R., 2014, Physical Review Letters, 113, 231102

\bibitem[{Rice(2014)}]{rice2014detection}
Rice K., 2014, Challenges, 5, 296

\bibitem[{Schulze-Makuch {et~al}\mbox{.}(2011)Schulze-Makuch, M{\'e}ndez,
  Fair{\'e}n, von Paris, Turse, Boyer, Davila, Ant{\'o}nio, Catling, \&
  Irwin}]{schulze2011two}
Schulze-Makuch D. {et~al.}, 2011, Astrobiology, 11, 1041

\bibitem[{Sellers {et~al}\mbox{.}(2013)Sellers, Margetts, Coria, \&
  Manning}]{sellers2013march}
Sellers W.~I., Margetts L., Coria R.~A., Manning P.~L., 2013, PloS one, 8,
  e78733

\bibitem[{{Silburt} {et~al}\mbox{.}(2015){Silburt}, {Gaidos}, \&
  {Wu}}]{2015SilburtPlanetRadii}
{Silburt} A., {Gaidos} E., {Wu} Y., 2015, ApJ, 799, 180

\bibitem[{Speakman(2005)}]{speakman2005body}
Speakman J.~R., 2005, Journal of Experimental Biology, 208, 1717

\bibitem[{Tarter {et~al}\mbox{.}(2011)Tarter, Ackermann, Barott, Backus, Davis,
  Dreher, Harp, Jordan, Kilsdonk, Shostak, {et~al.}}]{tarter2011first}
Tarter J. {et~al.}, 2011, Acta Astronautica, 68, 340

\bibitem[{Tegmark {et~al}\mbox{.}(2006)Tegmark, Aguirre, Rees, \&
  Wilczek}]{tegmark2006dimensionless}
Tegmark M., Aguirre A., Rees M.~J., Wilczek F., 2006, Physical Review D, 73,
  023505

\bibitem[{Tegmark \& Rees(1998)}]{tegmark1998cosmic}
Tegmark M., Rees M.~J., 1998, The Astrophysical Journal, 499, 526

\bibitem[{Vilenkin(1995)}]{PhysRevD.52.3365}
Vilenkin A., 1995, Phys. Rev. D, 52, 3365

\bibitem[{{Weinberg}(1987)}]{1987PhRvL..59.2607W}
{Weinberg} S., 1987, Physical Review Letters, 59, 2607

\end{thebibliography}

\end{document}